# Astro2020 Science White Paper
# High-Resolution Spectroscopic Surveys of Ultracool Dwarf Stars & Brown Dwarfs

**Thematic Areas:**  ☐ Planetary Systems  ☒ Star and Planet Formation
☐ Formation and Evolution of Compact Objects  ☐ Cosmology and Fundamental Physics
☒ Stars and Stellar Evolution  ☐ Resolved Stellar Populations and their Environments
☐ Galaxy Evolution  ☐ Multi-Messenger Astronomy and Astrophysics

**Principal Author:**
Name: Adam Burgasser
Institution: UC San Diego
Email: aburgasser@ucsd.edu
Phone: +1 858 822 6958

**Co-authors:**
Daniel Apai, University of Arizona
Daniella Bardalez Gagliuffi, American Museum of Natural History
Cullen Blake, University of Pennsylvania
Jonathan Gagne, University of Montreal
Quinn Konopacky, UC San Diego
Emily Martin, UC Santa Cruz
Stanimir Metchev, University of Western Ontario
Peter Plavchan, George Mason University
Ansgar Reiners, Georg-August-Universität Göttingen
Everett Schlawin, University of Arizona
Clara Sousa-Silva, MIT
Johanna Vos, American Museum of Natural History

**Abstract** (350 characters):
High resolution spectroscopy of the lowest-mass stars and brown dwarfs reveals their origins, multiplicity, compositions and physical properties, with implications for the star formation and chemical evolution history of the Milky Way. We motivate the need for high-resolution, infrared spectroscopic surveys to reach these faint sources.

The lowest-mass stars and brown dwarfs - collectively referred to as ultracool dwarfs [15] - encompass objects with masses ≤ 0.1 $M_\odot$, $T_{eff}$ ≤ 3000 K, and spectral types ≥ M7, including L, T and Y dwarfs. They are a highly abundant population, comprising at least 15% of all stars in the Milky Way [16], including many of the nearest systems to the Sun. They are also extremely long-lived; every ultracool dwarf ever formed exists today. Combined with their fully convective interiors and the time-dependent cooling of brown dwarfs, ultracool dwarfs have the potential to serve as useful probes of Milky Way star formation and chemical enrichment history, as well as benchmarks for models of cool atmospheres and thermal evolution.

High-resolution spectroscopy, defined here as $\lambda/\Delta\lambda$ ≥ 20,000 ($\Delta v$ ≤ 15 km/s), is a necessary tool for characterizing any stellar population in detail. Abundances, kinematics, rotation, multiplicity, magnetic field strength, spotting, and atmospheric dynamics all emerge from high-resolution spectroscopic studies. For intrinsically faint and red ultracool dwarfs, these investigations have been restricted to large aperture facilities (8-10m), with limited study by existing high-resolution spectroscopic surveys on 3-5m class telescopes (e.g., APOGEE [18], CARMENES [23], SPIROU [21]). With the implementation of red- and infrared-sensitive multi-object spectrographs (MOS) on large aperture facilities, and the development of advanced calibration and analysis methods, **the next decade will provide an opportunity to address outstanding science questions on the physical properties of ultracool dwarfs, and their use in studies of the Galactic environment.**

**Abundances & model atmospheres:** Mass and composition are the two most fundamental parameters dictating the structure and thermal evolution of stars. For main-sequence FGK stars, high-resolution spectroscopy has provided abundances for dozens of elements, which are used to map coeval populations, trace Galactic chemical evolution history, and assess the properties of orbiting exoplanets [5,27,33]. Abundances become particularly important for ultracool dwarfs (and giants), whose cool atmospheres are dominated by molecular opacity and are thus uniquely sensitive to abundance patterns, chemistry (including condensation and gas-grain chemistry), and atmospheric dynamics [6,9,31]. Moreover, the fully convective interiors of ultracool dwarfs enable bulk compositional measurements from the atmosphere (modulo condensate rainout; [17]), including interior compositional changes due to fusion, such as lithium and deuterium depletion [24,32]. In the absence of fusion, these nearly pristine probes can potentially trace chemical enrichment over the history of the Milky Way, including Big Bang nucleosynthetic abundances of lithium, boron, and other light elements [41]. Abundance ratios of stable elements, such as C/O, and isotopic ratios, are also critical tracers of origin, particularly at the (somewhat fuzzy) boundary between giant exoplanets and brown dwarfs [37,38].

Despite their scientific potential, abundances studies of ultracool dwarfs have been limited to modeling scaled solar abundance patterns [39,40], with some exceptions (Figure 1, [35,41]). Both the complexity and resource intensity of producing models with abundance variations has proven challenging. Retrieval methods [42] may provide a path forward for detailed abundance analysis, but only with concurrent theoretical investigation of abundance effects on gas

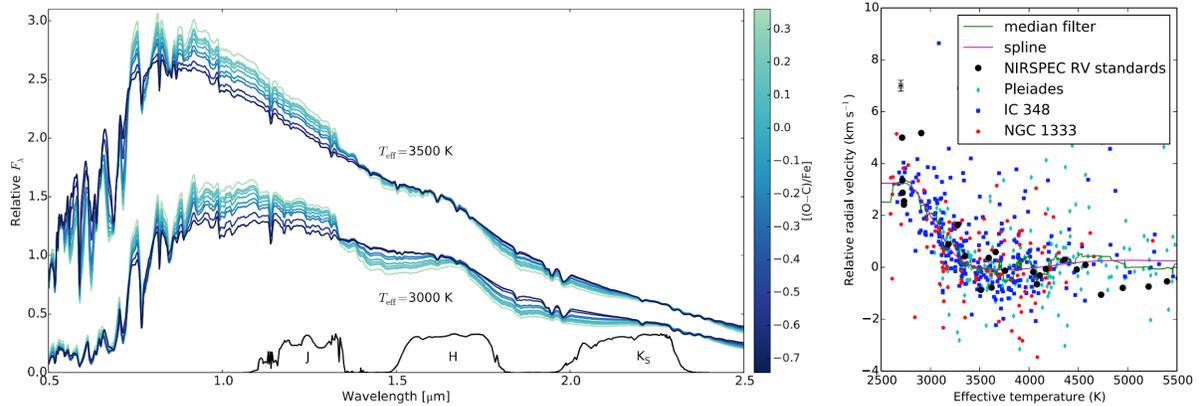

Figure 1: LEFT: C/O abundance variations in $T_{eff}$ = 3000 K (bottom) and 3500 K (top) stellar atmospheres. While the overall metallicity is unchanged, non-solar abundances can significantly influence the spectra of ultracool dwarfs, and is therefore a relevant parameter to constrain [35]. RIGHT: Radial velocities relative to cluster means for low-temperature members of the Pleiades, IC 348, and NGC 1333. Below $T_{eff}$ ≈ 3000 K, a consistent redshift of up to 3 km/s emerges. This is not likely physical, but instead highlights current errors in spectral modeling that can bias ultracool dwarf kinematic studies [11].

chemistry and condensation.

**Kinematics: Clusters:** Robust radial velocities (RVs) are required for studies of the lowest-mass stars and brown dwarfs in young clusters and associations, which serve as benchmarks for the evolution of temperature, luminosity, radius, rotation, activity, disks and planet formation [27,33,35]. Cluster membership often hinges on matching 6D Galactic phase coordinates (position and velocity) to the overall cluster, particularly for low-density, nearby associations [43]. Kinematics are also necessary to distinguish overlapping clusters, and can probe mass-dependent scattering, related to the age and dynamical state of the cluster; and cluster boundedness [44]. Studies using *Gaia* astrometry and ground-based RVs can probe these dynamics for stellar members, but the substellar population is barely reached. Future observations may clarify surprising trends, such as a temperature-dependent systematic redshift found among low-mass members of several nearby clusters, likely due to errors in the spectral models (Figure 1 [11]).

**Kinematics: Galactic Populations:** Kinematics distinguish major Galactic populations (e.g., thin disk, thick disk, halo) and trace the statistical age of a population. The latter is of particular interest for cooling brown dwarfs, which are expected to show systematic age variations with spectral type. Specifically, for a uniform, well-mixed field population, L dwarfs (a mix of low-mass stars and rapidly cooling brown dwarfs) have been predicted to be on average younger than both M dwarfs (mostly stars) and T dwarfs (mostly old brown dwarfs) [1,8]. Yet the observed 3D kinematics of local M and L dwarfs find the opposite trend [10,25,29], with some evidence of kinematic clumping. These samples are small (<100 sources) and subject to selection biases. Proper characterization of field ultracool dwarf ages, and the underlying mass and birth rate distributions, requires a high resolution spectroscopic survey to complement, e.g., *Gaia* astrometry.

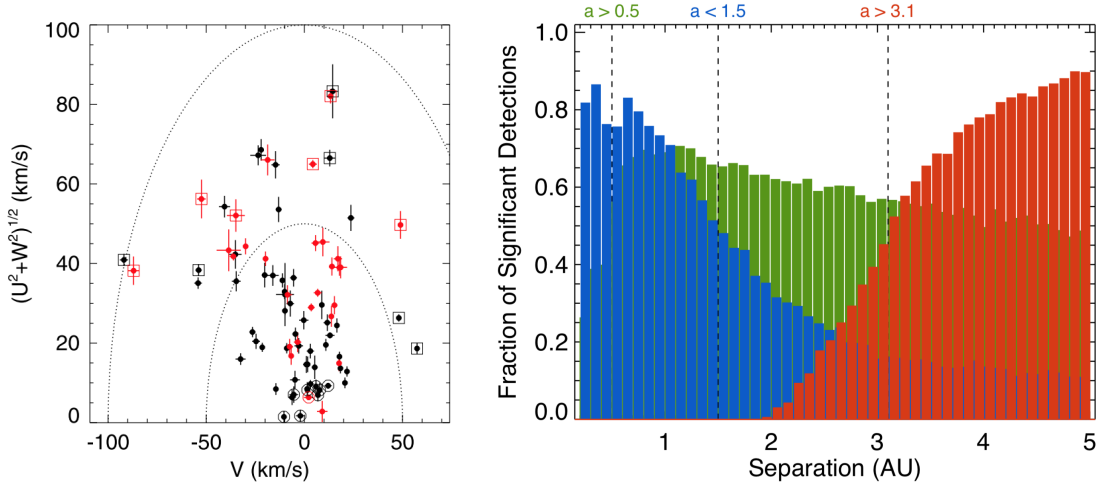

FIgure 2: LEFT: Toomre diagram for nearby late-M (black) and L dwarfs (red). The latter have a larger dispersion, contrary to population synthesis predictions [10]. RIGHT: Detection probabilities for an L+T dwarf binary at 20 pc based on RV monitoring (blue: 3 samples/year with $\sigma_{RV}$ = 0.5 km/s), astrometric monitoring (green; 8 samples/year with $\sigma_{AST}$ = 2 mas) and direct imaging (red: separation limit of 150 mas). RV variables sample the closest separations which provide mass and orbit measurements.

RV surveys would also discover "fly-by" stars, which have made (or will make) a close approach to the Sun. These systems may have a role in shaping the composition and orbits of the outer Solar System (including the hypothesized Planet 9), and the arrangement of the major Solar planets [22]. *Gaia* astrometry indicates a <1pc encounter rate of ≈20 stars/Myr, but only 15% of encounters within 5 pc and ±5 Myr having been identified due to the lack of RVs for ultracool dwarfs [2]. A pertinent example is WISE J0720-0846AB, a very low-mass star/brown dwarf binary whose large positive RV and near-zero proper motion indicate that it passed within 50,000 AU of the Sun 70,000 years ago [45]. The large number of T- and Y-type brown dwarfs in the Solar Neighborhood make them promising candidates for fly-by stars, whose identification will require high resolution spectroscopy to extremely faint infrared magnitudes ($M_J$ > 20).

**RV variables:** Multiplicity probes formation mechanisms and provides fundamental physical characteristics (masses, radii) for all stellar types. In the ultracool dwarf regime, the observational expense of RV monitoring of individual sources has limited detection of the most closely-separated multiples (Figure 2), which may comprise a significant fraction of all ultracool multiples [3]. A multi-epoch, sensitive infrared RV survey would provide a more effective mechanism for identifying large numbers of these systems, improving both multiplicity statistics (multiplicity fraction, distributions of separation, eccentricity, mass ratio) and identifying benchmarks for evolutionary and atmosphere models.

**Line Broadening: Angular Momentum Evolution and Magnetic Fields:** While FGK stars spin down over time due to magnetized winds [4,46], ultracool dwarfs span a transition from relatively rapid spindown (t < 1 Gyr) to effectively no spindown [13,36]. This is concurrent with a decline in optical and X-ray magnetic emission, attributed to increasingly neutral photospheres [20]. Yet magnetic (radio) emission persists in the coolest brown dwarfs [14], while there are few rotational vsini [47,48] or Zeeman broadening measures [49] in this regime to fully explore age-activity-rotation correlations. Both measures require high resolution infrared spectroscopy, as well as advancements in molecular opacities and Lande g coefficients [49].

Rotational velocities are a function of both age and mass, and there is evidence that young brown dwarfs extend the apparent spin-mass relation of Solar planets up to six orders of magnitude in mass, suggesting a "universal" law [50]. Such a remarkable trend requires validation with additional measurements, particularly for older, colder, and fainter brown dwarfs. With exceptional signal-to-noise (S/N > 100), it would also be possible to apply Doppler imaging methods to these systems to measure the surface structure of photospheric clouds [51] which drives photometric variability for nearly all brown dwarfs [52]. Crossfield [48] finds that 30m-class telescopes would enable surface maps for nearby L and T dwarfs, which can in turn be used to constrain 3D global climate models for brown dwarfs and exoplanets [53].

**Needs: Instrumentation & Facilities:** To achieve these and other science opportunities, it is necessary to conduct spectroscopic surveys that provide high infrared sensitivity and resolution, along with multiplexing in space and time. Single-pointing instrumentation on 30m class telescopes will be useful for individual targets of interest (e.g., ultracool dwarf exoplanet hosts, spectroscopic binaries), but novel science hinges on MOS capabilities. Instruments such as Subaru/PSF [54], VLT/MOONS [55], and TMT/MODHIS will be useful for cluster science outlined above. For the broader Galactic science, new high resolution spectroscopic survey facilities like the MaunaKea Spectroscopic Explorer [19] will be needed.

**Needs: Accurate molecular opacities for all relevant species.** Current experimental and theoretical molecular databases (ExoMol [34], HITRAN/HITEMP [12,28], TheoReTs [26]), contain hundreds of molecular species and isotopologues, but there remain dozens more important diatomic, triatomic and polyatomic species with inaccurate, incomplete, or absent opacities. Investment in these efforts support analyses of ultracool dwarfs, exoplanets, and many laboratory situations (e.g., flame gases).

**Needs: Adequate spectral parameters**. Correct estimation of spectral behaviors at the pressures, temperatures, and compositions of ultracool dwarfs, and in the presence of magnetic fields, requires detailed understanding of thermodynamic properties (e.g. partition functions, broadening coefficients) and splitting/shifting effects (e.g. Landé g-factors) for each relevant molecule. This is particularly true in strong-field environments, currently poorly modeled, as these are necessary to correctly infer surface magnetic field strengths to constrain the age-activity-rotation relation below the hydrogen fusion limit.